# Continuous topological evolution

**R. M. Kiehn**
Mazan, France

rkiehn2352@aol.com

http://www.cartan.pair.com

**Abstract**
A non-statistical theory of continuous, but irreversible, evolution can be constructed in terms of the Cartan calculus. The fundamental postulate, for an evolutionary theory which admits irreversible processes, is that the topology of the initial state will be different from the topology of the final state. Several fundamental theorems of continuous evolution are established, yielding a set of global conservation laws for reversible or irreversible processes. As examples, a comparison of the evolution of Topological Torsion and Topological Action is made for hydrodynamic and electromagnetic systems. The relationship between the evolution of Topological Torsion and a thermodynamically irreversible process is established.

Pacs 02.40.+m, 03.40.Gc, 03.65.Bz

## 1. Introduction

An objective of this article is to develop a theory of topological evolution that may be used to describe the irreversible evolution of dissipative non-conservative physical systems. The ideas will utilize topological concepts for it is postulated that a necessary condition for irreversible evolution involves topological change. The basis for such a postulate follows from the fact that if an evolutionary process is described by a map, $\Phi$, between initial and final states, and if the map is not continuously reversible, then the observable topology of the final state is different from the observable topology of the initial state. Cartan's methods can be used to extend these concepts to the dynamics of physical systems that admit description in terms of exterior differential forms. It is remarkable that the mathematical development leads to recognizable thermodynamic features which permit the determination of classes of processes which are reversible or irreversible. For example, all Hamiltonian processes are thermodynamically reversible. An essential feature of irreversible processes is that they involve the evolution of what has been defined as Topological Torsion.

The observation of topological change, with the production and destruction of defects and holes, lines of self-intersection and other obstructions, will be the signature of topological irreversible evolution. Topological change can occur discontinuously as in a cutting process, or continuously, as in a pasting process. Such *continuous* but irreversible processes can be used to study the decay of turbulence, but not its creation. The production of disconnected components will be the signature of those *discontinuous* processes which are necessary to describe the creation and evolution of chaotic but perhaps reversible evolution, or turbulent, irreversible evolution. In this article, emphasis will be place upon those processes which are continuous, but not reversible.

Processes or maps that preserve topology are technically described as homeomorphisms [1]. Homeomorphisms are both continuous and reversible. Homeomorphic reversibility means that the inverse function, $\Phi^{-1}$, must exist and must be continuous. Topological properties, such as orientability, compactness, connectivity, hole count, lines of self-intersection, pinch points, and Pfaff dimension are invariants of homeomorphisms, but geometrical properties such as size and shape are not necessarily invariants of homeomorphic deformations. In fact an elementary method of recognizing topological properties is to observe those properties that stay the same under continuous deformations.

The theory of Continuous Topological Evolution is developed herein in terms of physical systems that undergo certain thermodynamic processes. The physical system is assumed to be modeled in terms of the topological features inherent in Cartan's theory of exterior differential systems. The thermodynamic process will be defined in terms of a vector field, V, and its effect on the differential forms that make up the exterior differential system. The action of the process will be defined in terms of the Lie derivative with respect to V acting on the on the differential forms that make up the exterior differential system that define the physical system. The methods lead to concepts that are coordinate free and are well behaved in any reference system. A precise non-statistical definition of topological irreversibility will be stated, and a cohomological equivalent of the first law of thermodynamics will be derived and studied relative to the single constraint of continuous but irreversible topological evolution. Remarkably, many intuitive thermodynamic concepts can be stated precisely, without the use of statistics, in terms of the theory of continuous topological evolution based on the Cartan topology.

Given a topology on the final state and a map from an initial state to the final state it is always possible to define a topology on the initial state such that the given transformation, or even a given set of transformations, is continuous. However, the topologies of the initial and final states need not be the same; hence the map need not be reversible. Recall that with respect to a discrete topology all maps from the initial to final state are continuous, while relative to the concrete topology, only the constant functions are continuous [2]. A first problem of a theory of topological evolution is to devise a rule for constructing a topology that is physically useful and yet is neither too coarse nor too fine. Such a rule is necessary for the concept of continuity of an evolutionary transformation is defined relative to the topologies of the initial and final states. In this article the topological rules will be made by the specification of an exterior differential system that will model the physical system of interest. Many physical systems appear to be adequately modeled by 1-form of Action.

Physical exhibitions of continuous and discontinuous transformations can be achieved through the deformations of a soap film attached to a wire frame. For example, a soap film attached to a single closed, but double, loop of wire can be deformed from a non-orientable surface into an orientable surface continuously. (The topological property of orientability is changed) That is, the soap film can be transformed from a Mobius band into a cylindrical strip. As another example, consider an initial state where a soap film is attached to two slightly separated but concentric circular wire loops. The resulting surface is a minimal surface of a single component. As the separation of the concentric rings is slowly increased, the minimal surface is stretched until a critical separation is reached. Then, without further displacement, the surface spontaneously continues to deform to form a cone. The surface separates at the conical singularity, and the two separate sheets of the cone continue to collapse to form a minimal surface of two components. The final state consists of two flat films attached, one each, to each ring. The originally connected minimal surface undergoes a topological (phase) change to where it becomes two disconnected (still minimal) surfaces. An example of this topological transition in the surface of null helicity density has been described in conjunction with the parametric saddle node Hopf bifurcation of a Navier-Stokes flow [3].

In this article the fundamental set, X, will be the points {x,y,z,t,...} that make up an N-dimensional space. Upon this fundamental set will be constructed arbitrary subsets, such as functions, tensor fields and differential forms. Many different topologies may be constructed on the fundamental set in terms of special classes of subsets that obey certain rules of logical closure. In fact the very existence of subsets can be used to define a course topology on X in terms of a topological base. The topological base consists of those subsets whose unions form a special collection of all possible subsets that is closed under logical union and intersection. This special collection of subsets will be defined as the open sets of a topology. The topological base can be used to define a topological structure. A space is said to have a topological structure if it is possible to determine if a transformation on the space is continuous [4].

## 2. Continuity

The classic definition [5] of a continuous transformation between a set X with topology T1 to a set Y with a topology T2 states that the transformation is continuous if and only if the inverse image of open sets of T2 are open sets of T1. This definition can be made transparent by use of a simple point set example. Consider two sets of 4 points, an initial state, {a,b,c,d} and a final state {x,y,z,t}. Define an open set topology on the initial state T1 = [X,∅,a, ab,abc] and a open set topology on the final state T2 = [X,∅,x,y,xy,yzt]. The transformation considered is exemplified by the Figure 1.

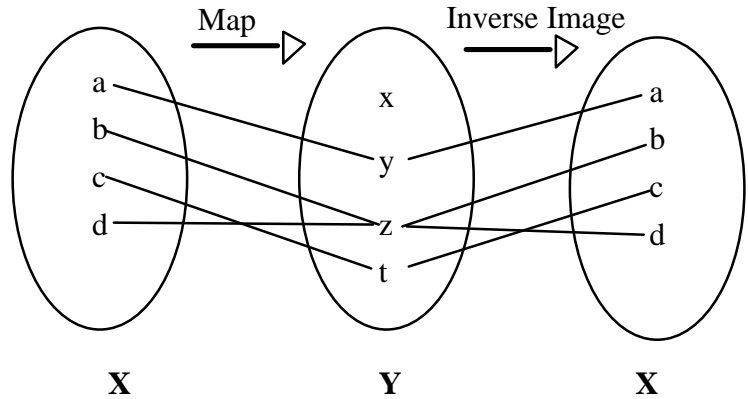
Figure 1.

The open set (y) has a preimage (a) which is open. The open set (yzt) has a preimage (abc) which is open. Hence the Map is continuous. (The open sets that involve x are not included as the map does involve x.) However, the Inverse Image mapping is not continuous for the open set (ab) has a preimage as (yz) but (yz) is not an open set of Y. The point set example demonstrates the idea of a continuous but not homeomorphic mapping. The objective herein is to examine such maps in terms of exterior differential systems.

There exists another more useful method of defining continuity which does not depend explicitly on being able to define open sets and their inverse images. This second method of defining continuity is based on the concept of closure. The closure of a set can be defined in (at least) two ways:

The closure of a set is the union of the interior and the boundary of the set.
The closure of a set is the union of the set and its limit points.

The first definition of closure is perhaps the most common, and is often exploited in geometric situations, where a metric has been defined and a boundary can be computed. The second definition of closure is independent from metric and is the method of choice in this article both for defining continuity and establishing a topological structure. In terms of the concept of closure, a transformation is continuous if and only if for every subset, the image of the closure of the initial subset is included in the closure of the image of that subset [5]. Another way of stating this idea is: a map is continuous iff the limit points of every subset in the domain permute into the closure of the subsets in the range.

If a method for constructing a closure operator ( a Kuratowski closure operator K of a subset relative to a topology) can be defined, then a strong version of continuity would imply that the Kuratowski closure operator commutes with those transformations which are continuous. The test for continuity would be to construct the closure of an arbitrary subset on the initial state, and then to propagate the elements of the closure to the final state by means of a transformation. If this result is the same as the result obtained by first propagating the subset to the final state by means of the transformation, and then constructing its closure on the final state, then the map is continuous. Note that such a procedure has defined a topological structure which will be exploited in this article, for the subsets of interest will be defined as a Cartan system {Σ} of exterior differential forms on X. The topological base defined by this class of sets is too course to be of interest. The Cartan exterior derivative will be used to generate additional sets of forms, {dΣ}, which when adjoined to the initial system of forms defines the Kuratowski closure as the system of forms, K(Σ) = {Σ ∪ dΣ).

The Cartan exterior product may be used as a convenient intersection operator between sets of differential forms. Starting from the system, {Σ}, the Cartan topology is then determined by the construction of the Cartan -Pfaff sequence, which consists of all possible intersections that may be constructed from the subsets of the closure of the differential system,

$$\text{the Pfaff Sequence}$$
$$\{\Sigma, d\Sigma, \Sigma^\wedge d\Sigma, d\Sigma^\wedge d\Sigma ...\}. \qquad (2.1)$$

For example if Σ is a single 1-form of Action, then the Pfaff sequence consists of n ≤ N terms which are non-zero. The number n is defined as the Pfaff dimension or class of the 1-form. The subsets of the Cartan topological space consist of all possible unions of the subsets that make up the Pfaff sequence. The Cartan topology will be constructed from a topological basis which consists of the odd elements of the Pfaff sequence, and their closures,

$$\text{The Cartan topological base :}$$
$$\{\Sigma, K(\Sigma), \Sigma^\wedge d\Sigma, K(\Sigma^\wedge d\Sigma)...\} \, . \qquad (2.2)$$

With respect to a topological base constructed from a single 1-form of Action it has been shown [34] that the Cartan exterior derivative may be viewed as a closure or limit point operator. Given any subset of the Cartan topological space, the exterior derivative of that subset generates its limit points, if any. This is a remarkable result, for as will be demonstrated below, all $C^2$ vector fields acting through the concept of the Lie derivative on a set of differential forms, with $C^2$ coefficients, generate continuous transformations with respect to the Cartan topology. Moreover, the Cartan topology is disconnected if Σ^dΣ is not zero.

### 3. The Evolutionary Process

A simple evolutionary process, X ⇒ Y, is defined by a map Φ. The map, Φ, may be viewed as a propagator that takes the initial state, X, into the final state, Y. In this article the evolutionary processes to be studied are asserted to be generated by vector fields of flow, V. The vector fields need not be topologically constrained such that they are generators of a single paramenter group. In other words, kinematics without fluctuations is not imposed a priori. The trajectories defined by the vector fields may be viewed as propagators that carry domains into ranges in the manner of a convective fluid flow. The evolutionary propagator of interest to this article is the Lie derivative with respect to a vector field ,V, acting on differential forms, Σ [6]. The Lie derivative has a number of interesting and useful properties.

1. The Lie derivative does not depend upon a metric or a connection.

2. The Lie derivative has a simple action on differential forms producing a resultant form that is decomposed into a transversal and an exact part:

$$L(V) w = i(V)dw + d\, i(V)w. \qquad (3.1)$$

3. The Lie derivative may be used to describe deformations and topological evolution.

In contrast to the covariant derivative, the Lie derivative may be used to describe deformations and shear processes associated with convective fluid flow, and processes for which the topology changes continuously. The covariant derivative has many almost equivalent definitions, but it is designed such that it is an operation of displacement which convert tensors into tensors. Hence it is a differential procedure constrained to admit diffeomorphisms, and therefore not useful to describe topological evolution. In certain cases it is restricted to generate processes that preserve the line element therefore is useful only to the study of isometries (rigid body motions, and bending without shear). The covariant derivative is avoided in this article

For example, the action of the Lie derivative on a 0-form (scalar function) is the same as the directional derivative of ordinary calculus,

$$L(V)\,\varphi = i(V)d\varphi + 0 = V \bullet \operatorname{grad} \varphi. \qquad (3.2)$$

In the examples given below, it will be demonstrated that the action of the Lie derivative on a 1-form typically will generate hydrodynamic or electromagnetic equations of motion. The Lie derivative is not the same as the classic covariant derivative, or its generalizations used in gauge or fiber bundle theories. The reason is that the Lie derivative satisfies the equations

$$L(fV)\,\omega = f\, L(V)\,\omega + df \wedge i(V)\omega. \qquad (3.3)$$

The covariant derivative and its generalizations are constrained such that the second term on the right vanishes:

$$D(fV)\,\omega = f\, D(V)\,\omega\,. \qquad (3.4)$$

This equation is often interpreted by saying that f represents the action of the "group", and the covariant derivative is defined such that it commutes with the action of the group. The Lie derivative is not limited to the constraint of a specified group.

4. With respect to vector fields and forms constructed over $C^2$ functions, the Lie derivative commutes with the Kuratowski closure operator. Hence, the Lie derivative generates transformations on differential forms which are continuous with respect to the Cartan topology.

The first three ideas appear in the literature, but the extraordinary property that all $C^2$ vector fields that propagate $C^2$ differential forms in the manner of a convective flow (Lie derivative) are continuous relative to the Cartan topology requires proof: Given $\Sigma$, first construct the closure, $\{\Sigma \cup d\Sigma\}$. Next propagate $\Sigma$ and $d\Sigma$ by means of the Lie derivative to produce the decremental forms, say Q and Z,

$$L(V) \Sigma = Q \quad \text{and} \quad L(V) d\Sigma = Z. \qquad (3.5)$$

Now compute the contributions to the closure of the final state as given by $\{Q \cup dQ\}$. If $Z = dQ$, then the closure of the initial state is propagated into the closure of the final state, and the evolutionary process defined by V is continuous. However,

$$dQ = d\ L(V)\Sigma = di(V)d\Sigma + dd(i(V)\Sigma) = di(V)d\Sigma$$
as $dd(i(V)\Sigma = 0$ for $C^2$ functions. But,
$$L(V)d\Sigma = d\ (i(V)d\Sigma) + i(V)dd\Sigma \quad = di(V)d\Sigma = Z$$
as $i(V)dd\Sigma = 0$ for $C^2$ functions.

It follows that $Z - dQ = 0$ for $C^2$ functions and vector fields, V, and therefore subject to the constraint of $C^2$ differentiability, every vector field, V, generates a continuous evolutionary process relative to the Cartan topology.   QED

The Lie derivative also can be used to make some sense out of certain classes of discontinuous evolutionary processes (which are not $C^2$). For example, consider a vector field $V = \rho v$ where the support function, $\rho$, is not $C^2$. Then the action of the Lie derivative produces the discontinuity or excess function,

$$Z - dQ = d\rho \wedge i(v)\Sigma + \rho\ dd(i(v)\Sigma). \qquad (3.6)$$

This equation is of use in the study of shock waves and other discontinuous processes. Note that special considerations arise when $i(v)\Sigma = 0$. Such vector fields are defined to be associated vector fields, while those vector fields that satisfy $i(v)d\Sigma = 0$ are defined, for reasons that will become apparent, as extremal vector fields relative to $\Sigma$.

### 4. Topological Evolution

If the flow field generated by V acting on a Cartan system of forms satisfies the equations
$$L(V)\Sigma = 0 \quad \text{and} \qquad (4.1)$$
$$L(V)d\Sigma = 0, \qquad (4.2)$$

then the forms of the closure are said to be absolute invariants of the flow. It follows that each element that makes up the Cartan topological base is also invariant, such that the whole Cartan topology is invariant. As V is continuous, and the topology is preserved, those vector fields, V, that satisfy the equations above must be homeomorphisms and are reversible. In other words, $Q = 0$ and $dQ = 0$ are sufficient conditions that V be reversible for $C^2$ systems.

However, for continuous transformations on the elements of the C2 Cartan topology the general equations of topological evolution become,

$$L(V)\Sigma = Q \quad \text{and} \tag{4.3}$$
$$L(V)d\Sigma = dQ, \tag{4.4}$$

from which it follows that

$$L(V)(\Sigma \wedge d\Sigma) = Q \wedge d\Sigma + \Sigma \wedge dQ \quad \text{and} \tag{4.5}$$
$$L(V)(d\Sigma \wedge d\Sigma) = 2\, dQ \wedge d\Sigma. \tag{4.6}$$

As these equations of continuous topological evolution imply that the elements of the topological base may not be constant, then specific tests must be made to determine what features of the topology are changing, if any. For if it can be determined that the topology is indeed modified by the evolutionary process, then the process generated by this class of vector fields, V, is continuous, but not reversible.

When $dQ \neq 0$, the limit points are not constants, and it would be natural to expect that the topology is not constant. However, even if Q is closed, such that $dQ = 0$, it may be true that Q contains harmonic components, such that deRham classes of the Action are not constant. Again the conclusion is reached that such continuous flows are not reversible, for the hole count of the initial state is not the same as the hole count of the final state.

## 5. Simple Systems

For purposes of expose, the Cartan system, $\Sigma$, will be limited to a single 1-form of action, A., and perhaps a single pseudoscalar field, or N form density, $\rho$. The 1-form of Action, A can be written in several equivalent formats:

$$A = A_\mu dx^\mu = \mathbf{p} \bullet d\mathbf{x} - H dt = L dt + \mathbf{p} \bullet (d\mathbf{x} - \mathbf{v} dt) \tag{5.1}$$

The last representation indicates that the Action may be viewed abstractly in terms of the kinematic fluctuations in position,

$$\Delta \mathbf{x} = (d\mathbf{x} - \mathbf{v} dt), \tag{5.2}$$

on a fluctuation space of 10 dimensions {t,**x**,**v**,**p**}. The coefficients **p** act as Lagrange multipliers for the fluctuations. The constraint of canonical momentum, $\partial L/\partial \mathbf{v} = \mathbf{p}$, reduces the dimensionality to a state space of 7 dimensions, and the further constraint of kinematic perfection $\Delta \mathbf{x} = (d\mathbf{x} - \mathbf{v} dt) = 0$ reduces the dimensionality to the configuration space of 4 dimensions. These concepts will be exploited in the examples given below.

The exterior derivative of A produces a 2-form of closure points, F = dA, whose components are given by the expression, $F_{\mu\nu} dx^\mu \wedge dx^\nu$. The combined set { A, F} forms the closure of the set A. All possible intersections of the closure, { A, F, A^F, F^F }, form the Pfaff sequence for the domain {x,y,z,t}, with elements:

| | |
|---|---|
| TOPOLOGICAL ACTION | $A = A_\mu dx^\mu$ |
| TOPOLOGICAL VORTICITY | $F = dA = F_{\mu\nu} dx^\mu dx^\nu$ |
| TOPOLOGICAL TORSION | $H = A \wedge dA = H_{\mu\nu\rho} dx^\mu dx^\nu dx^\rho$ |
| TOPOLOGICAL PARITY | $K = dA \wedge dA = K_{\mu\nu\rho\sigma} dx^\mu dx^\nu dx^\rho dx^\sigma.$ |

The union of all elements of the Pfaff sequence and their closures forms the elements of the Cartan topological base:

$$\{ A, A \cup F, H, H \cup K \}. \qquad (5.3)$$

In order to take into account projective (and certain discontinuous) features, the vector fields of interest will be scaled by a support function, $\rho$, such that $J = \rho V$. The fundamental equations of *continuous* evolution become

$$L(\rho V)A = Q \qquad (5.4)$$
$$L(\rho V)F = dQ \qquad (5.5)$$
$$L(\rho V)H = Q \wedge F + A \wedge dQ \qquad (5.6)$$
$$L(\rho V)K = 2\, dQ \wedge F = 2\, d(dQ \wedge A) = d(Q \wedge F) \qquad (5.7)$$

Note that for the even dimensional elements of the Pfaff sequence, (F and K) the action of the Lie derivative is to produce an exact form, $dQ$ for eq.(5.5) and $2\, d(Q \wedge F)$ for eq. (5.7). As integrals of exact forms over closed cycles or boundaries of support vanish, then it is possible to formulate the first theorem (Poincare).

**Theorem I:**   *All even dimensional Pfaff classes of p-forms, $dA = F$, $dA \wedge dA = K$, ... are relative integral invariants of continuous evolutionary processes relative to the Cartan topology.*

The closed integrals of F, K, ... are invariants of the flow as $L(\rho V) \oint F = \oint dQ = 0$, for closed integration cycles of support. It should also be noted that the domains of support of the even dimensional Pfaff classes are *not* compact without boundary.

### 6. Cohomology and the Evolution of Energy
#### 6.1 Cartan's Magic Formula and the first law.

Consider the first equation (5.4) of the set, and write, using Cartan's Magic formula,

$$L(\rho V)A = i(\rho V)dA + d(i(\rho V)A = Q. \qquad (6.1)$$

Define $i(\rho V)A$ as the function U and $i(\rho V)dA$ as the inexact 1-form W,

$$i(\rho V)dA = W, \quad i(\rho V)A = U. \qquad (6.2)$$

Then equation (5.4) becomes equivalent to the statement of cohomology; that is, the difference between the inexact 1-form Q and the inexact 1-form W is equal to a perfect differential for continuous evolution:

$$L(\rho V)A = W + dU = Q. \tag{6.3}$$

Cartan's Magic formula, expressing the propagation of the 1-form of Action down the tube of trajectories generated by the vector field V, becomes the dynamical equivalent of the first law of thermodynamics, when the inexact 1-forms Q and W are interpreted as heat and work respectively. Examples will show that this is no accident. Fundamentally, equation (6.3) is a topological law describing the evolution of energy. It is remarkable that the first law follows, without axiomatization, from the single and simple constraint that the 1-form of action, A undergoes continuous evolution. It is also intuitively pleasing to see that the inexact 1-forms, Q and W, are defined in terms of a process. Elementary discussions of heat and work often emphasize the energy content of the first law, rather than the engineering idea that heat and work are related to processes. Further examples given below will demonstrate that equation (6.3) is indeed related to the evolution of energy.

### 6.2 Thermodynamic Irreversibility and the Torsion vector.

It is important to realize that Q represents the inexact 1-form of heat, and its integral is the measurable quantity. When $L(V)A = Q = 0$ then the topological evolution process is defined to be an adiabatic process. When $L(V)Q = R \neq 0$, the process is defined to be radiative. When $Q \wedge dQ \neq 0$, then the heat 1-form is said to be non-integrable. The implication is that there does not exist an integrating factor for Q. However, classical thermodynamics states that a process that creates a heat 1-form which does not admit an integrating factor is thermodynamically irreversible. Hence, given a physical system described in terms of a 1-form of Action, A, it is possible for a given process to compute Q and dQ. If $Q \wedge dQ = 0$, then that process is thermodynamically irreversible. It is also possible to solve for those processes V that are thermodynamically irreversible when applied to a specified physical system..

Given an Action 1-form in 4D, construct the 3 form of Topological Torsion, $H = A \wedge dA$. Then there exists a vector field T such that $i(T) dx \wedge dy \wedge dz \wedge dt = A \wedge dA$. This vector field is defined as the Topological Torsion vector. The properties of the Topological Torsion vector are such that

$$i(T)dA = \Gamma\, A, \text{ and } i(T)A = 0. \tag{6.4}$$

Hence evolution of the physical system defined by A, in the direction of T, yields the formula

$$L(T)A = \Gamma\, A = Q, \text{ and } L(T)dA = d\Gamma \wedge A + \Gamma\, dA = dQ. \tag{6.5}$$

The heat 3-form becomes

$$Q \wedge dQ = L(T)A \wedge L(T)dA = \Gamma^2\, A \wedge dA \tag{6.6}$$

If the physical system admits Topological torsion, then $A \wedge dA$ is not zero. Hence if the coefficient $\Gamma^2$ is not zero then the Heat 3-form is not integrable and the process is thermodynamically irreversible.

Examples will show that Γ is equal to the coefficient of the Topological Parity 4 form, dA^dA. When dA^dA is not zero, such that Γ is not zero, the Torsion vector is uniquely defined, for then the coefficients of the 2-form F = dA form an anti-symmetric matrix with an inverse. The physical system is said to define a symplectic 4D manifold. The conclusion is that thermodynamic irreversibility is an artifact of 4 dimensions [42].

### 6.4  Other thermodynamical properties

Other authors have emphasized the topological foundations of thermodynamics [7], and from the time of Caratheodory have noted the connection to Pfaff systems [8]. However, these authors did not have access to, or did not utilize, the Cartan topology and DeRham cohomology. A remark by Tisza, "... the main content of thermostatic phase theory is to derive the topological properties of the sets of singular points in Gibbs phase space" [9], greatly stimulated the early developments of the theory presented in this article.

As must be the case in thermodynamics, there is a fundamental difference between the 1-form W and the 1-form Q. From the definition i(ρV)dA = W , it follows that

$$i(\rho V)i(\rho V)dA = i(\rho V)W = 0 \quad \text{(transversality)} . \quad (6.7)$$

This fact implies that the 1-form W must be constructed from first integrals, φ, of the flow V, or from transversal fluctuations in the kinematics:

$$W = d\phi + \mathbf{f} \bullet (d\mathbf{x} - \mathbf{v}dt). \quad (6.8)$$

Although W can be included in the concept of Q, there are parts of Q that are not transformable into W. The 1-form of Work, W, is always transversal, i(ρV)W = 0, but it is not necessarily true that Q be transversal in the sense that i(ρV)Q = 0. This issue is at the heart of the second law of thermodynamics. The argument is pleasing for it gives formal substance to the intuitive differences between the thermo-dynamic concepts of heat and work.

All continuous processes may be put into equivalence classes as determined by the vector fields, V, that generate the flow. For example, for any p-form, ω, those vector fields that satisfy the transversal equation,

$$\text{Associated} \quad i(\rho V)\omega = 0, \quad (6.9)$$

are said to be elements of the associated class of vector fields relative to the form ω. Those vectors that satisfy the equations,

$$\text{Extremal} \quad i(\rho V)d\omega = 0, \quad (6.10)$$

are said to be elements of the extremal class of vector fields. It should be noted that the 2-form dω admits a unique extremal vector only on spaces of odd dimensions (2n+1) dimensional state space called a contact manifold). If the Pfaff dimension of the 1-form A is 4, then extremal vectors do not exist. The domain is a symplectic manifold of even dimension  However, on the symplectic manifold it then follows that there exists a unique Torsion vector.

Vectors which are both extremal and associated are said to be elements of the characteristic class of vector fields [10].

$$\text{Characteristic} \quad i(\rho V)\omega = 0, \text{ and } i(\rho V)d\omega = 0, \quad (6.11)$$

Note that characteristic flow lines generated by V of the Characteristic class preserve topology, for each form of the Cartan topological base is invariant with respect to the action of the Lie derivative relative to characteristic flows.

Characteristics are often associated with wave phenomena. In the next section, the continuous vector fields that generate evolutionary processes will be put into two distinct categories, and several subcategories that have both physical and topological significance. The associated class of vector fields will have particular significance to the study of certain discontinuous processes.

It is important to realize that Q represents the inexact 1-form of heat, and its integral is the measurable quantity. When $L(V)A = Q = 0$ then the topological evolution process is defined to be an adiabatic process. When $L(V)Q = R \neq 0$, the process is defined to be radiative. When $Q \wedge dQ \neq 0$, then the heat 1-form is said to be non-integrable. The implication is that there does not exist an integrating factor for Q. However, classical thermodynamics states that a process that creates a heat 1-form which does not admit an integrating factor is thermodynamically irreversible. Hence, given a physical system described in terms of a 1-form of Action, A, it is possible for a given process to compute Q and dQ. If $Q \wedge dQ = 0$, then that process is thermodynamically irreversible. It is also possible to solve for those processes V that are thermodynamically irreversible. Examples of this idea and its expression in terms of the Topological Torsion 3-form will be given below.

### 7. Closed flows

Continuous flows or processes are naturally divided into two main categories: those for which $dQ \neq 0$ (open flows) and those for which $dQ = 0$ (closed flows). Closed flows also will be defined as uniformly continuous flows, to distinguish them from open flows, which are also continuous relative to the C2 constraint:

$$\text{Closed flow} \quad L(\rho V) \, dA = 0 \quad (7.1)$$

defines a uniformly continuous closed flow, while

$$\text{Open flow} \quad L(\rho V) dA = dQ \neq 0 \quad (7.2)$$

defines an open flow. Flow in the direction of the Torsion vector is an open flow.

Uniform continuity implies that the limit sets are invariant. Continuity only requires that the limit points permute amongst themselves. For example a fold into pleats which are then pasted together is a processes that rearranges the limit points and is not therefor uniformly continuous. Hence uniform continuity is a more constrained situation. When $dQ = 0$, it is possible to formulate immediately the following theorem (Poincare) for closed flows:

> **Theorem II:** *All even dimensional Pfaff classes of p-forms, $dA = F$, $dA \wedge dA = K$, ... are invariants of closed evolutionary processes that satisfy $dQ = 0$ relative to the Cartan topology. The forms F, K, ... form a set of absolute integral invariants with respect to closed, or uniformly continuous, flows.*

The proof of the theorem follows immediately from (7.1) and by application of the Leibnitz rule:
$$L(\rho V)\{dA \wedge dA \wedge ... \wedge dA\} =$$
$$(L(\rho V)dA) \wedge dA \wedge ... \wedge dA + dA \wedge (L(\rho V)dA) \wedge ... \wedge dA + ... \quad = 0. \tag{7.3}$$
Also note that
$$L(\rho V)\int dA = \int L(\rho V)dA = 0 \quad \text{(Invariance of open integrals)}. \quad \text{Q.E.D}$$

The first application of theorem II gives,

$$L(\rho V)dA = L(\rho V)F = 0, \tag{7.4}$$

which is the equivalent of Helmholtz' theorem [11] and the conservation of angular momentum per unit moment of inertia, or the conservation of Topological Vorticity.

The second application of theorem II gives:

$$L(\rho V) \, dA \wedge dA = L(\rho V)K = 0, \tag{7.5}$$

which leads to the conservation of Topological Parity, with respect to closed flows. In general, the formula

$$L(\rho V)dA \wedge dA \wedge ... dA = 0, \tag{7.6}$$

expresses the invariance of a 2N dimensional area with respect to continuous flows.

These results are to be compared with the even dimensional Poincare absolute integral invariants [12] for the more restrictive case of Hamiltonian (extremal) evolution of a Hamiltonian action, $A = A_\mu dx^\mu = p \bullet dx - H dt$, on a 2N+1 dimensional state space. It is the result (7.6) which is interpreted in statistical mechanics as the invariant area of phase space with respect to extremal, or Hamiltonian, evolution. The fact of the matter is that uniform continuity alone (Equation 7.1) produces a set of absolute integral invariants for any action, in Hamiltonian format or not. Hamiltonian extremal flows satisfy equation (6.10) and are therefore uniformly continuous, but they are not the only flows that satisfy equation (7.1). The invariance of "phase space area" is a consequence of uniform continuity alone, and does not require the additional constraints of constant homogeneity that limit the set of continuous flows to that subset of continuous vector fields which are extremal, and Hamiltonian.

Consider the third equation (5.6) of the set, and write

$$L(\rho V) H = i(\rho V)dH + d(i(\rho V)H) = Q \wedge F, \qquad (7.7)$$

for the evolution of the topological torsion 3-form with respect to closed flows. Define the 3-form $J = i(\rho V)dH$, and recall that J must be transversal to V. Therefore, J must be a current 3-form with the format,

$$J = i(\rho V)dH = \rho(dx - V^x dt) \wedge (dy - V^y dt) \wedge (dz - v^z dt). \qquad (7.8)$$

For closed flows, $dJ = d(L(\rho V)H) = L(\rho V)K = 0$, hence the closure criteria implies that

$$dJ = \{div \rho V + \partial \rho / \partial t\} dx \wedge dy \wedge dz \wedge dt = 0. \qquad (7.9)$$

In classical physics, when the term in brackets is set to zero the resulting equation is described as the "equation of continuity", and is taken to be an axiom of most problems in classical hydrodynamics. A more precise statement is that the null bracket represents the "conservation of mass", for the 4-form, $M = \rho\, dx \wedge dy \wedge dz \wedge dt$ is an invariant of the uniformly continuous flow, $L(V)M = 0$.

Again a precise topological statement clarifies the intuitive thermodynamic idea that closed systems do not exchange mass with their surroundings. The concept is even a bit more general, for if the flow is not closed, in the sense that $dQ \neq 0$, but parity is conserved, $dQ \wedge F = 0$, then (7.9) is still true and it follows that there exists a mass conservation law. The important idea that the conservation of topological parity is to be associated with the conservation of mass serves to focus attention on the importance of the 3-form of topological torsion in evolutionary systems. Note that the exact part of (7.7), $d(i(\rho V)H)$, is also a current, but it is always conserved in the sense of an electromagnetic charge-current conservation law, as $dd(i(\rho V)H) = 0$.

### 8. DeRham categories of Closed Vector Fields

DeRham's cohomology theory [13] may be used to classify p-forms, and such ideas may be applied to the 1-form W defined by equation (6.1-6.3)). Correspondingly, the vector fields that are used to construct the 1-forms W can be put into the following categories depending on whether W is null, exact, closed, or not closed with respect to exterior differentiation. These categories are defined as:

**Closed Flows with a 1st law format:  Q-W =dU**
 HAMILTONIAN CATEGORY :   $i(\rho V)F = W = 0$    $dQ=dW=0$
 EULER-BERNOULLI CATEGORY :  $i(\rho V)F = W = d\phi$   $dQ=dW=0$
 STOKES CATEGORY  :  $i(\rho V)F = W = d\phi + \gamma$  $dQ=dW=0$
**Open flows**
 NAVIER STOKES FLOWS :  $i(\rho V)F = W$      $dQ=dW \neq 0$.

  The functions, $\phi$, must be first integrals as in general, $i(V)W = 0$.  For closed flows the first law (6.3) insures that the 1-form W is closed, $dW = dQ = 0$, but W need not be exact and may contain harmonic components, $\gamma$.  That is, the 1-form W is not necessarily representable over the variety $\{x,y,z,t\}$ in terms of the gradient of a single scalar function.  The classic example of a non-exact 1-form is given by the expression,

$$\gamma = \sigma_z \, (ydx - xdy)/(x^2+y^2) \tag{8.1}$$

for which $d\gamma = 0$, but the integral over a closed cycle is $\oint \gamma = 2\pi\sigma_z$.  The coefficient $\sigma_z$ assumed to be a constant.

  Such forms, $\gamma$, generate period integrals and the DeRham cohomology classes.  The number of independent forms of the type given by equation (8.1) determine the Betti numbers of a connected variety for which the singular point (at the origin in the example) has been excised.  The Betti numbers can be interpreted as a method for counting the number of holes or handles in the variety.  It is these contributions to the general differential form that carry topological information about the domain of support.  The duals to these forms are also closed, leading to the definition, *harmonic* forms.

  From the first law (6.3) the harmonic contributions to W are equal to the harmonic contributions to Q.  If the harmonic contributions to Q are not zero, then the number of "holes and handles" in the Cartan topology of the final state is different from the number of holes and handles in the Cartan topology of the initial state, and the evolutionary process is continuous but not reversible in the sense that the inverse image exists and is continuous.

  In order to make (8.1) transversal, use the Cartan trick of substituting $dx^i - V^i dt$ for each $dx^i$.  The transversal harmonic form becomes

$$\gamma = \sigma_z \, (ydx - xdy)/(x^2+y^2) + \sigma_z \, (\mathbf{r} \times \mathbf{V})_z dt/(x^2+y^2) \tag{8.2}$$

which demonstrates the close relationship to transversal harmonic forms and angular momentum.  The format may be extended to a spin vector of components

$$\sigma = [\sigma_z/(x^2+y^2),\ \sigma_x/(y^2+z^2),\ \sigma_y/(z^2+x^2)] \tag{8.3}$$

such that the harmonic form becomes

$$\Gamma = \sigma_z(ydx - xdy) + \sigma_y(xdz - zdx) + \sigma_x(zdy - xdz) + (\boldsymbol{\sigma} \cdot \mathbf{r} \times \mathbf{V})dt . \quad (8.4)$$

and the last term becomes a type of "spin orbit" coupling term. The idea of harmonic contributions to a 1-form is closely related to the concept of a complex number or ordered pair representation; i.e., the form cannot be represented by a map to a space of 1 dimension. Other formats for harmonic 1-forms are:

$$\Gamma = \{\phi\, d\chi - \chi d\phi\}/ \{\pm \phi^2 \pm \chi^2\} \quad (8.4)$$

where $\phi$ and $\chi$ are arbitrary functions, and

$$\Gamma = \{\phi\, d\phi^* - \phi^* d\phi\}/ \{\phi^*\phi\}. \quad (8.5)$$

The last representation of a harmonic form is in the format of the "probability current" of quantum mechanics, and gives a clue as how to adapt the formalism of this article to quantum systems. Such a development is deferred to a later article.

For closed flows on space time, the fundamental equations of evolution are given by the expressions for the odd 1-form and the odd 3-form. The even forms are invariant. The two fundamental equations of uniformly continuous evolution are:

$$L(\rho V)A = Q \quad (8.6)$$
$$L(\rho V)H = Q\wedge F. \quad (8.7)$$

It should be remarked that if the 1-form of Action, A, is completely integrable in the sense of Frobenius, then the 3-form H is evanescent, and the evolutionary equation (8.7) has no applicability. Such evolutionary processes are the equivalent to laminar flows in fluid dynamics and completely integrable, non-chaotic, Hamiltonian systems. It is known that if a Lagrangian system is not chaotic, then the action is reducible to two variables (or less), and the 3-form H is necessarily zero. However when there exists a sense of helicity in the evolutionary process, or chaos is present, then (8.7) describes the appropriate topological evolution.

The first expression (8.6) may be put into correspondence with the evolution of action, while the second fundamental equation (8.7) may be described as the evolution of complexity, or perhaps better as the evolution of defects, links, knots, or in abstract terms, the evolution of an entropic concept. If the heat 1-form Q is zero, then the evolutionary process is adiabatic, and topology is preserved. However, as the Cartan topology is not connected when H ≠ 0, then continuous evolution of H can be accomplished only between connected subsets. The transition from a connected topology with H = 0 to a disconnected topology with H ≠ 0 can only take place via a discontinuous transformation. The idea is that the *continuous* rate of change of H is definite (and arbitrarily taken to be positive). This feature is one of the key properties of entropy. Entropy can never change its sign. The *creation* of topological torsion, H, is a discontinuous process, but its growth (or decay) can be described by a continuous process (relative to the Cartan topology). These entropic features of the topological torsion 3-form will be useful in the description of the transition to turbulence.

### 9. The Hamiltonian Sub-Category

It should be remarked, that Cartan has proved that if

$$i(\rho V)F = W = 0, \quad Q = dU \qquad (9.1)$$

for any reparametrization, $\rho$, then V generates a Hamiltonian system, and visa versa [14]. This remarkable result indicates that Hamiltonian flows are not only continuous, but preserve many topological properties. The 1-form Q must be exact for Hamiltonian flows. Hence the observable holes and handles are topological invariants of Hamiltonian flows, as the $\Gamma$ terms vanish. However, the fact that Q is exact for Hamiltonian flows does not completely establish a proof that Hamiltonian systems preserve all topological properties of the Cartan topology.

In the calculus of variations, vector fields that satisfy (6.10) are defined as extremal vector fields. Characteristic vector fields are a subclass of extremal fields that satisfy the equations

$$L(\rho V)A = 0 \quad \text{and} \quad L(\rho V)F = 0. \qquad (9.2)$$

In other words, continuous characteristics preserve the Cartan topology (Q = 0 and dQ = 0). Characteristic Hamiltonian vector fields generate waves in systems that can be endowed with the additional structure of a metric.

### 10. The Euler - Bernoulli subcategory

The Euler – Bernoulli category is not quite Hamiltonian in the extremal sense. W is not zero, but must be a perfect differential, $W = d\phi$. However, this perfect differential must be a first integral in order to satisfy the transversality condition, $i(\rho V)W = 0$. The 1 form Q is not necessarily so constrained. The abstract flows of this category are to be compared with the equations of motion of a compressible Eulerian fluid in which there may be stratification. If the pressure, P, is a function of the density, $\rho$, alone, then the Eulerian flow can be reduced to a Hamiltonian system [15]. If there exists some anisotropy due to stratification, then the Hamiltonian reduction is not perfect. Note that the first integral, $\phi$, acts as a Bernoulli constant along a given streamline, but the constant can vary from streamline to streamline because the function is transversal.

### 11. The Stokes subcategory

The Stokes category admits topological evolution in the sense that the harmonic contributions to W are not null, and therefore the "hole and handle" count of the Cartan topology is changing in an evolutionary manner. Such closed flows are not reversible. Note that all closed flows preserve topological vorticity and topological parity, and so if the flow is without vorticity in the initial state, then the flow is without vorticity in the final state. The Pfaff dimension [16] remains less than 2. However, if the initial state has vorticity, that vorticity will be preserved, but the Topological Torsion 3-form can change. In fact the Topological Torsion 3-form could be non-zero in the initial state, and zero in the final state, for the decay rate of topological torsion is proportional to $Q^\wedge F$. Both the 1-form of action and its hole count, and the 3-form of Topological Torsion, and its twisted handle count, are not necessarily invariants of a Stokes flow.

A method of distinguishing between "holes and twisted handles" is of some interest. Note that physically a handle can be constructed by deforming the rims of two holes in a surface into tubes and pasting the tubular ends together. If the rims are twisted by half integer or integer multiples of pi before the ends are glued together, then the handles have torsion. Note that a handle cannot be constructed in the plane, so it is an intrinsically 3-dimensional thing. If the 3-form H vanishes, then there are no handles in the initial state, and as the Hamiltonian evolution produces no more new holes, there can be no more new handles in a Hamiltonian flow. However, existing handles may become twisted or knotted, because $Q^\wedge F \neq 0$, even for Hamiltonian flows. These facts correspond to the physical result that Hamiltonian systems are not dissipative and preserve energy, but that does not mean that entropy must be conserved. for all closed flows, $dW = 0$ and the transversality condition $i(\rho V)W = 0$ implies that the form W is an absolute invariant of the flow (the enthalpy):

$$L(\rho V) W = 0 \qquad \text{closed flows} \qquad (11.1)$$

### 12. The Navier-Stokes category of open flows

It should be noted that the 1-form Q may be use to construct the Pfaff sequence, $\{Q, dQ, Q^\wedge dQ, dQ^\wedge dQ\}$, and generates another Pfaff dimension depending upon the rank or class of the elements of the Pfaff sequence for Q. For closed flows, $dQ = 0$ and the Pfaff dimension generated by Q is 1. For open flows, $dQ \neq 0$, but the Pfaff sequence demonstrates that the topological features of open flows can have various levels of complexity. For example, the criteria that the Pfaff dimension of Q be 2 or less is equivalent to the Frobenius integrability constraint, $Q^\wedge dQ = 0$. This is precisely the Caratheodory condition that there exist "inaccessible paths" [17], and that (on a simply connected neighborhood) the 1-form of heat be representable as, $Q = TdS$. The criteria implies that the associated process is not thermodynamically irreversible. The topological evolution theory presented herein permits an analysis to be made for non-equilibrium processes, where the heat 1-form is not of the equilibrium monomial format, $Q = TdS$. For the Navier-Stokes flow, the key feature is that $dQ = dW \neq 0$, but it still must be true that W is transversal. Therefore the 1-form W must be constructed from fluctuations, in the format ,

$$W = \mathbf{f} \bullet (d\mathbf{x} - \mathbf{v}dt) \quad + \text{ closed additions transversal to V.} \qquad (12.1)$$

For open flows with a Lagrange multiplier f, W is no longer a flow invariant. In the examples below (section 17), a particular choice is made for f which will generate the Navier-Stokes equations, which may have equilibrium or non-equilibrium solutions.

### 13. The Topological Base in terms of an Action 1-form

For continuous evolution in space-time, the key idea is that the exterior differential system consists of a Pfaff sequence constructed from a single 1-form of Action A, plus (perhaps) some additional constraints defining a domain of support and its boundary. The work of Arnold (and others) [18] has established that the singular points (zero's) of a global 1-form carry topological information. This idea is to be extended to the singular points of all elements of the Pfaff sequence, or topological base. In Appendix A, the idea of how a global 1-form of Action, A, existing on a space of dimension N+1 can be put into correspondence with a line bundle on a variety of dimension N is worked out in detail. The key features are that the Jacobian matrix of the projectivized 1-form of Action carries most of the information about the subspace. The similarity invariants of the Jacobian matrix determine the curvature properties of the subspace. The anti-symmetric components of the Jacobian are the functions that make up the projectivized 2-form, F = dA. The polynomial powers of F form the Chern classes for the line bundle.

For continuous transformations on a variety of {x,y,z,t}, the Cartan Action, A, can be defined in a kinematic sense as:

$$A = \mathbf{v} \bullet d\mathbf{r} - H\, dt \qquad (13.1)$$

where the "Hamiltonian" function, $H$, is defined as,

$$H = \mathbf{v} \bullet \mathbf{v}/2 + \int dP/\mathbf{r}. \qquad (13.2)$$

Substitute this 1-form onto the constraint of Hamiltonian evolution, $i(\mathbf{v},1)dA = 0$. Carry out the indicated operations of exterior differentiation and exterior multiplication to yield a system of necessary partial differential equations yields of the form,

$$\partial\mathbf{v}/\partial t + \operatorname{grad}(\mathbf{v} \bullet \mathbf{v}/2) - \mathbf{v} \times \operatorname{curl} \mathbf{v} = -\operatorname{grad} P/\rho \qquad (13.3)$$

Equations (13.3) are exactly the Euler equations for the evolution of a perfect fluid. By direct computation, the 2-form F = dA of topological vorticity has components,

$$F = dA = \omega_z dx \wedge dy + \omega_x dy \wedge dz + \omega_y dz \wedge dx + a_x\, dx \wedge dt + a_y\, dy \wedge dt + a_z\, dz \wedge dt, \qquad (13.4)$$

where by definition

$$\boldsymbol{\omega} = \operatorname{curl} \mathbf{v}, \qquad \mathbf{a} = -\partial\mathbf{v}/\partial t - \operatorname{grad} H. \qquad (13.5)$$

These vector fields always satisfy the Poincare-Maxwell-Faraday induction equations, dF = ddA = 0 for C2 functions, or,

$$\operatorname{curl} \mathbf{a} + \partial\boldsymbol{\omega}/\partial t = 0,\ \operatorname{div} \boldsymbol{\omega} = 0. \qquad (13.6)$$

The 3-form of Helicity or Topological Torsion 3-form, H, is constructed from the exterior product of A and dA as,

$$H = A \wedge dA = H_{ijk}\, dx^i \wedge dx^j \wedge dx^k$$
$$= -T_x dy \wedge dz \wedge dt - T_y dz \wedge dx \wedge dt - T_z dx \wedge dy \wedge dt + h\, dx \wedge dy \wedge dz, \qquad (13.7)$$

where **T** is the fluidic Torsion axial vector current, and h is the torsion (helicity) density:

$$\mathbf{T} = \mathbf{a} \times \mathbf{v} + H\boldsymbol{\omega}, \tag{13.8}$$
$$h = \mathbf{v} \bullet \boldsymbol{\omega} \tag{13.9}$$

The Torsion current, **T**, consists of two parts. The first term represents the shear of translational accelerations, and the second part represents the shear of rotational accelerations. The topological torsion tensor, $H_{ijk}$, is a third rank completely anti-symmetric covariant tensor field, with four components on the variety {x,y,z,t}. Note that the helicity density is the fourth component of this third rank tensor field that transforms covariantly under all diffeomorphisms, including a Galilean translation.

The Topological Parity becomes

$$K = dH = dA \wedge dA = 2(\mathbf{a} \bullet \boldsymbol{\omega}) dx \wedge dy \wedge dz \wedge dt \tag{13.10}$$

The fundamental law for the local evolution of this set is given by the expression

$$\text{div } \mathbf{T} + \partial h/\partial t = -2(\mathbf{a} \bullet \boldsymbol{\omega}), \tag{13.11}$$

and yields the helicity-torsion current conservation law if the anomaly, $-2(\mathbf{a} \bullet \boldsymbol{\omega})$, on the RHS vanishes. It follows that the integral of H over a boundary of support vanishes by Stokes theorem. This idea is the generalization of the conservation of the integral of helicity density in an Eulerian flow. Note the result is independent from viscosity, subject to the constraint of $K = 0$

The torsion vector, **T**, consists of two parts. The first term represents the shear of translational accelerations, and the second part represents the shear of rotational accelerations. The pseudo scalar function, K, acts as the source for the divergence of the torsion vector, **T**, and the torsion or helicity density, h. When $K = 0$, the evolutionary "lines" associated with the torsion tensor never cross, implying that the system is free of defects in space time. If K is positive or negative, the defects in the system are either growing or decaying. Equation (13.11) is the fundamental new law of topological physics that governs the specific realizations of controlled processes that minimize or maximize defect evolution.

Recall that if $H = A \wedge dA = 0$, the 1-form of action satisfies the complete integrability condition of Frobenius. Similar to the Caratheodory equilibrium result for Q, the flow can be described then in terms of two variables; i.e., the flow is laminar. Turbulent flow is not laminar, and the transition from the laminar to the turbulent state must involve the topological evolution of H. It was the evolution of the 3-form of topological torsion that galvanized the author's interest in topological evolution. The 3-form, H, and its evolution are intuitively related to the thermodynamic property of entropy. The fact that the Cartan topology is disconnected if the topological torsion, H, is not zero implies that the turbulent state cannot be created from the laminar state by means of a continuous transformation. Turbulence must be created by a discontinuous process. However, the decay of turbulence can be described by means of continuous process.

### 14. Global Conservation Laws (first variation)

Extremal (or Hamiltonian) flows and Eulerian flows induce a set of global conservation laws in the sense that the closed integrals of all odd dimensional Pfaff classes of the fundamental forms are relative integral invariants of uniformly continuous evolution. The result follows from the fact that the evolutionary rates, Q and Q^F are exact with respect to such flows. Integrals of exact forms evaluated over closed cycles, whether the cycle ( Z1 or Z3) is a boundary or not, vanish. Hence all closed integrals of odd dimensional sets, $\oint$ A and $\oiiint$ H, are evolutionary invariants of Hamiltonian and Eulerian flows.

For the closed flows of the Stokes category, the evolutionary rates of all odd Pfaff classes are closed, but not necessarily exact. That is,

$$dQ = 0, \text{ and } d(Q^\wedge F) = 0, \qquad (14.1)$$

implying closure, but Q and Q^F are not exact. The DeRham classes are not empty and are not necessarily flow invariants. Topology changes during such evolutionary processes.

Hence a global set of conservation laws in terms of closed integrals of A and H can be devised only for those closed chains that satisfy Stokes theorem, and those chains must be boundaries (of support). Arbitrary closed integrals are not evolutionary invariants. This lack of relative integral invariance [19] for $\oiiint$ H corresponds to the production or destruction of 3 dimensional defects, and these new defects are indications of changing topology and changing inhomogeneity. Formally, a closed integral over a closed form is a period integral whose value, by Brouwer's theorem [20], is an integer multiple of some smallest value. A variation of a period integral signals a change in a Betti number and hence a change in topology. Such flows can produce three dimensional defects.

These results point out the limitations of Moffatt's and Gaffet's claims [21] that the volume integral of helicity density, **v** • curl **v**, is an evolutionary invariant. Helicity is NOT necessarily an invariant of a continuous flow. Moreover, open or closed integrals of Helicity are not necessarily integral invariants of continuous evolution. In particular, the closed volume integral of helicity density, the fourth component of the Helicity four current, is not an invariant of continuous flows for which there is a torsion current . However, a theorem depending on only the first variation can be stated for the continuous evolution of flows restricted to Hamiltonian or Eulerian flows:

> **Theorem III:** *The (uniformly) continuous evolution of all odd dimensional Pfaff classes of the Cartan base with respect to Hamiltonian or Eulerian flows (dQ = 0, Q exact) are exact. Hence, the closed integrals of A and H = A^dA over closed cycles or boundaries are relative integral invariants with respect to Hamiltonian or Eulerian flows.*

The proof of the theorem is as follows:

$$L(\rho V)A = i(\rho V)dA + d(i(\rho V)A) = d[P + i(\rho V)A] = Q \text{ and is exact.}$$
$$\text{Therefore } L(\rho V)\oint A = \oint Q = 0 \Rightarrow \text{invariance of } \oint A.$$

Similarly,
$$L(\rho V) H = L(\rho V)A\wedge F = Q\wedge F = d[(P + i(\rho V) F], \text{ is exact such that}$$
$$L(\rho V) \oiiint H = \oiiint d(A\wedge F) = 0 = \text{invariance of } \oiiint H. \quad \text{Q.E.D.}$$

In the hydrodynamic case of a compressible Eulerian fluid, this theorem is the generalization of the "invariance of Helicity theorem" often stated for a barotropic domain or isentropic constraints. Closed flows therefore exhibit global conservation laws based on relative integral invariants of A and H, as well as absolute integral invariants of F and K. As will be demonstrated below, the integral of the 3-form of topological torsion, not the helicity density, over a boundary is an invariant of all flows that satisfy the Navier-Stokes equations and for which the vorticity vector field satisfies the Frobenius complete integrability conditions. This result is independent from the magnitude of the viscosity coefficient. On the other hand, the continuous destruction of 3-dimensional defects can be associated with closed flows of the Stokes category. Helicity is NOT necessarily a relative integral invariant of Stokes flows. Remarkably, such flows also admit a set of relative integral invariants, but these are determined only in terms of a second variational process.

### 15. Global Conservation Laws (Second Variation)

t should be noted that the *second* Lie derivative of the odd dimensional Pfaff classes (represented by A and H) does produce a set of global conservation laws for uniformly continuous processes. The result follows from the fact that the second Lie derivative of the Action with respect to closed flows is exact, where the first Lie derivative is closed! The fundamental theorem is then:

> **Theorem IV:** *The (uniformly) continuous evolution of all odd dimensional Pfaff classes of the Cartan base with respect to closed flows (dQ = 0) are closed, but not necessarily exact. The second Lie derivative is always exact so that $\oint Q$ and $\oiiint Q\wedge F$ are relative integral invariants of (uniformly) continuous (dQ = 0) evolution.*

The proof of the fundamental theorem is as follows:
$$L(\rho V) A = i(\rho V)dA + d(i(\rho V)A) = Q$$
$$L(\rho V) L(\rho V)A = L(\rho V) Q = R$$
$$= i(\rho V)d(i(\rho V)dA + di(\rho V)di(\rho V)A$$
$$= 0 \text{ (as } dQ = 0) + d\Gamma$$
$$= d\Gamma, \text{ which is exact.}$$

Similarly,
$$L(\rho V)L(\rho V)A\wedge dA = L(\rho V)Q\wedge F = d(\Gamma\wedge dA),$$
which is exact. It follows that
$$L(\rho V)\oiiint Q\wedge F = \oiiint d(\Gamma\wedge dA) = 0$$
such that $\oiiint Q\wedge F$ is a relative integral invariant. Q.E.D.

Uniform continuity requires that d (L(ρV) A) = L(ρV)dA = dQ = 0, which insures that Q and Q^F are closed. Hence closed integrals of the odd dimensional p-forms of Q and Q^F (and not necessarily A and H) are relative integral invariants of uniformly continuous evolution. The integrals $\oint Q$ and $\oiiint Q$^F generate global conservation laws for uniformly continuous processes in which dQ = 0. In elementary terms, on a space time variety, the fundamental theorem of uniformly continuous evolution states that the Lorentz force has zero curl, and the torsion defect production rate has zero divergence (K = 0), whether the system is dissipative or not.

The successive Lie derivation with respect to a *uniformly continuous* vector field J = ρV produces an exact sequence, starting from the concept of action-angular momentum, A, evolving to a *closed* set, Q, which under continued Lie derivation evolves to an *exact* kernel of radiation-power, R [20]. A similar exact sequence can be constructed for all odd dimensional Pfaff classes, A, A^dA, A^dA^dA, ....

## 16. Continuity and the Integers

A most remarkable feature of the fundamental theorem of uniformly continuous evolution is that the integral of any radiation 1-form, R, through a container which is a maximal cycle is in relation to the integers. This concept is another application of the Brouwer degree of a map theorem, that says that all period integrals are integer multiples of some smallest value. The maximal cycle is a closed set that is not a boundary but can contain a system with internal defects, hence the name, the "container". As a simple example consider a disc with several internal holes; the maximal cycle is the cycle which would be the boundary if the disc had no holes. The global conservation laws stated above imply that radiation through the maximal cycle must be compensated by a change in the cohomology class, or the production of a defect of inhomogeneity in the interior. Radiation defects ("holes and torsion handles") are quantized, for it is impossible to create half a hole.

It would appear from the above argument that Planck's hypothesis of quantized radiation oscillators may be considered a consequence of theorem IV and *Uniformly CONTINUOUS* evolution as defined by equation (7.1).

## 17. The Navier-Stokes Fluid

It is possible to gain insight into the viscous compressible evolution of A Navier – Stokes fluid in terms of open flow (dQ ≠ 0). The evolution equations are then constrained by the full format given by equations (5..4-5.7). The kinematic topology is often too course for direct application to a typical physical system. Additional topological constraints must be applied. For a Navier-Stokes fluid, the additional topological constraints on the admissible flow fields, V = {**v**, 1} implies a specific format is required for the dissipative force, f, given in (12.1); let f take the form μ curl ω such that upon dividing through by ρ, (12.1) becomes:

$$W = i(V)dA = -\{(\upsilon \; \text{curl}\omega)_i \; (dx^i - v^i dt)\}. \qquad (17.1)$$

Evaluating both sides explicitly and comparing coefficients of the terms $dx^i$ yields the Navier-Stokes partial differential equations,

$$\partial v/\partial t + \text{grad}(\mathbf{v} \bullet \mathbf{v}/2) - \mathbf{v} \times \text{curl } \mathbf{v} = -\text{grad } P/\rho + \upsilon \text{ curl curl } \mathbf{v} \qquad (17.2)$$

This process is typical of the Cartan method, where by the coefficients of a system of differential forms are equivalent to a system of partial differential equations. For the kinematic Action, equation (17.2) is the differential form equivalent to the Navier-Stokes equations; the constraint limits the class of all V to those V that are solutions to the Navier-Stokes partial differential equations.

The constraint given by (17.1) may be used evaluate the behavior of the topological base with respect to the evolution described by V. For example, the evolution of the Action is given by the expression,

$$L(V) A = i(V)dA + d\{i(V)A\} = -\{(\upsilon \text{ curl } \omega)_i ( dx^i - V^i dt)\} + d\{(\mathbf{v} \bullet \mathbf{v}/2) + H\} \qquad (17.3)$$

The evolution of the limit sets is given by

$$L(V) dA = -d\{(\upsilon \text{ curl } \omega)_i ( dx^i - V^i dt)\}. \qquad (17.4)$$

If the flow V is uniformly continuous, then the RHS of (17.4) must vanish, making $F = dA$ a flow invariant. The Navier-Stokes equations have $C^2$ solutions that belong to the Stokes category of closed flows. This result is an extension of the Helmholtz theorem on the conservation of vorticity. It would follow that the 4-form, $K = dA \wedge dA$ is also a flow invariant, for uniformly continuous flows. A remarkable result is that even for dissipative Navier Stokes flows where $\upsilon$ curl $\omega \neq 0$, it is still possible that the RHS of (17.4) may vanish, and the flow is uniformly continuous. Examples of such harmonic solutions to the Navier Stokes equations were presented by this author at the Permb conference on Large Scale Structures [3]. One such harmonic closed form solution was shown to develop a tertiary Hopf bifurcation in terms of the parameter of mean flow. The surface of null helicity density, $h = \mathbf{v} \bullet \omega = 0$ went through a topological phase change as the bifurcation took place similar to that which can be observed in a soap film supported between two rings is gradually extended in the axial direction.

According to theorem II, the even dimensional topological properties {F , K} are invariants of a uniformly continuous flow. If topology is to change in a uniformly continuous manner, the only possible candidates for topological evolution must be the 1-dimensional circulation, A, and the 3-dimensional torsion, H. For incompressible flows (div $\mathbf{v} = 0$) circulation defects must be associated with boundaries; however, if $K \neq 0$, then according to (13.11) torsion defects can occur within the bulk media. It is the author's perception that the production of torsion defects is the key to the understanding of large scale structures in continuous media, and the transition to turbulence [36].
In general, as has been stated above, if the flow is continuous, then the limit sets $d\Sigma$ must remain within the closure of $\Sigma$. Abstractly this idea can be written as,

$$L(V) \, d\Sigma = d\Sigma + \Sigma^\wedge \Sigma \tag{17.5}$$

Uniform continuity is the stronger constraint,

$$L(V) \, d\Sigma = 0. \tag{17.6}$$

For the Navier-Stokes flows, where the evolution is not necessarily uniformly continuous, the Navier-Stokes constraint (17.11) may be used to express the acceleration term, a, dynamically; i.e.,

$$\mathbf{a} = - \text{grad } H - \partial \mathbf{v}/\partial t = - \mathbf{v} \times \text{curl } \mathbf{v} + \upsilon \{ \text{curl curl } \mathbf{v} \} \tag{17.7}$$

By substituting this expression for **a** into equation (13.8) a simple engineering representation is obtained for the torsion vector current, T, of a Navier-Stokes fluid:

$$\mathbf{T} = \{ h \mathbf{v} - L \text{ curl } \mathbf{v} \} - \upsilon \{ \mathbf{v} \times (\text{curl curl } \mathbf{v}) \} \tag{17.8}$$

Note that the torsion axial vector current persists even for Euler flows, where $\upsilon = 0$. When $h = 0$, the torsion axial vector is proportional to the vorticity of the flow. It is the opinion of this author that many of the visual phenomena of fluid dynamics which have been associated with "vortices" are actually representations of torsion defects. In fact, a closed form solution to the Navier-Stokes equations was presented at the Permb conference [3] which indicates that the experimental phenomena of "vortex" bursting can be emulated by the streamlines of a flow for which there is no parametric evolutionary change of vorticity, but for which there is a parametric evolution and topological phase change of the 3-form of topological torsion. As the critical value of flow is achieved, a re-entrant compact torsion bubble is produced in what was originally a unidirectional flow.

The measurement of the components of the Torsion vector have been completely ignored by experimentalists (and theorists) in hydrodynamics (and other dynamical systems). By a similar substitution using the value of a given by the constraint (39), the topological parity pseudo-scalar becomes expressible in terms of engineering quantities as, $2 \upsilon ( \boldsymbol{\omega} \bullet \mathbf{curl } \boldsymbol{\omega})$ The measurement of the topological parity pseudo-scalar also have been completely ignored by experimentalists in hydrodynamics. The Topological Parity 4-form can be evaluated by exterior differentiation as,

$$K = dH = dA^\wedge dA = - 2 \upsilon ( \boldsymbol{\omega} \bullet \mathbf{curl } \boldsymbol{\omega}) \, dx^\wedge dy^\wedge dz^\wedge dt. \tag{17.9}$$

From this expression it is apparent that if the vorticity field is integrable in the sense of Frobenius, then viscosity does NOT contribute to the creation of torsion defects. As described below, the integral of K over {x,y,z,t} is related to the Euler index induced by the flow on the space time variety. If K = 0, the flow lines never intersect. The methods of this section have been used to describe coherent structures in hydrodynamic situations (including turbulent flows) [37] and to demonstrate that continuous time dependent 2 dimensional turbulence is a myth [38].

### 18. Pfaff's Problem, Characteristics, and the Torsion Current.

Closely related to the concept of topological torsion is the Pfaff problem that asks about the solubility of the system of differential equations defined by setting each element of the Cartan closure to zero. The problem is equivalent to finding characteristic vector fields which, if continuous, generate an evolutionary flow that preserves the Cartan topology. The key idea of Pfaff's problem is to find maps from spaces of q dimensions into the variety, X, such that when these maps and their differentials are substituted into the system of forms that make up the Cartan closure, then the new forms are equal to zero. In this sense, the pullback of the forms of the Cartan closure to the spaces of dimension q are zero. In the case of usual interest to physics, the maps are of a single parameter which almost always is associated with the concept of time. However, there may exist higher dimensional solutions of say two parameters or more.

The question arises as to the largest dimension of such a "solution" and is determined in terms of the "characters" and "genus" of the Pfaff system [22]. It is the objective of this section to demonstrate that the genus of the Pfaff system built from a single 1-form of action is 3 if the Torsion current, T, vanishes, and can be 2 only if $T \neq 0$. The genus is an arithmetic invariant and a topological property. A change of genus implies topological evolution. However for the special Pfaff system described, the characters are such that only 1-parameter solutions are possible, when $T = 0$, and a unique 2 parameter solution is admissible only when $T \neq 0$. In other words the Pfaff problem admits a "string" solution (a two parameter solution) only when the Torsion current is not zero.

Consider an electromagnetic format. For the electromagnetic case, the Cartan 1-form may be defined in terms of the vector and scalar potentials,

$$A = \mathbf{A} \bullet d\mathbf{r} - \phi dt. \tag{18.1}$$

Using the classical notation of Sommerfeld, define the E and B field intensities as

$$\mathbf{B} = \text{curl } \mathbf{A}, \quad \mathbf{E} = -\partial \mathbf{A}/\partial t - \text{grad } \phi. \tag{18.2}$$

Then the components of the Darboux-Cartan-Maxwell field, $F_{\mu\nu}$, may be written as an anti-symmetric matrix ( or as a Sommerfeld six-vector) of components :

$$\begin{array}{ll} F12 = Bz & F14 = Ex \\ F13 = -By & F24 = Ey \\ F23 = Bx & F34 = Ez \end{array}$$

The components of $dA = F$ become,
$$F = F_{\mu\nu} dx^\mu dx^\nu \tag{18.3}$$

The Topological torsion, H, becomes

$$H = A \wedge dA = -\{ \mathbf{E} \times \mathbf{A} + \phi\mathbf{B}, \mathbf{A} \bullet \mathbf{B}\}_{\mu\nu\sigma} dx^\mu dx^\nu dx^\sigma. \tag{18.4}$$

with the torsion current defined as,

$$\mathbf{T} = \mathbf{E} \times \mathbf{A} + \phi\mathbf{B} \tag{18.5}$$

and the helicity density,

$$h = \mathbf{A} \bullet \mathbf{B}. \tag{18.6}$$

The Topological Parity 4-form becomes the global top Pfaffian on the 4 dimensional space-time variety, and is equal to

$$K = dA \wedge dA = -2 \mathbf{E} \bullet \mathbf{B}\, dx \wedge dy \wedge dz \wedge dt. \tag{18.7}$$

It is of interest to show that propagation in the direction of the Torsion vector yields the formula

$$L(T)A = i(T)dA + d(i(T)A) = (\mathbf{E} \bullet \mathbf{B}) A + 0 = \Gamma A, \tag{18.8}$$

such that the dissipation coefficient is related to the 4 divergence of the Torsion current, $d(A \wedge dA) = dA \wedge dA$. Note that div $\mathbf{T} + \partial h/\partial t = -2\mathbf{E} \bullet \mathbf{B}$. The 3-form of axial current, H, is NOT conserved when $K \neq 0$. This result has been observed by Berger [23]. As demonstrated in section 6, evolution in the direction of the Torsion vector is thermodynamically irreversible.

Now the Pfaff problem is determined by the equations $A = 0$, $F = 0$. Following Slebodzinsky, as there is only one 1-form in the Pfaff system, the first character, s0, of the Pfaff system is equal to 1. Multiply F by $\varphi$, and use $A = 0$ to eliminate $\varphi dt$ in the equation $F = 0$. The result is given by the equation,

$$\{ \mathbf{E} \times \mathbf{A} + \varphi\mathbf{B} \}_{\mu\nu} dx^\mu dx^\nu = T_{\mu\nu} dx^\mu dx^\nu = 0, \tag{18.9}$$

which is an expression that does not contain dt. The polar system of these resultant equations determines the genus of the Pfaff system. In particular, if T, the torsion current vanishes, then (54) vanishes, the second character, s1 is zero and the genus of the Pfaff system is 3. All higher characters vanish, so the Pfaff system is special. Only 1-parameter homeomorphic evolutionary solutions are possible for the Pfaff system in 4 dimensions, when $\mathbf{T} = 0$.

On the other hand, for any arbitrary vector field, V, such that the two 1-forms

$$\{\mathbf{T} \times \mathbf{V} \}_\mu dx^\mu \text{ and } A, \tag{18.10}$$

are linearly independent, then the second character, s1, equals 1, and the genus is 2. There then exists a two parameter characteristic evolutionary system (a string). In other words, the presence of the torsion current is necessary for the existence of a two parameter solution to the Pfaff problem. There are no 3 parameter solutions to this Pfaff problem in 4-dimensions. This extraordinary connection between the concept of the Torsion current and the solubility of Pfaff's problem serves to further emphasize the content of the often neglected quantity of topological torsion.

The methods described in this section have been extended recently to explore the topological evolution of the electromagnetic fields [39] and the photon spin and other topological features of electromagnetism [40]. The Exterior differential system is bases upon the 1-form A and a globally closed N-1 form J. The method permits the utilization of another 3-form (the Topological Spin, A^G) which was discovered in 1969 [41] as well as the 3-form of Topological Torsion.

**19. The Euler index**
The coefficients of the Action 1-form globally define a covariant vector field on the variety. This vector field need not be a section without singularities. As mentioned in section 13 Arnold has shown how the singular points (zeros) of the Action 1-form, A, can be used to define the Euler index of the topology induced on the variety. Another method for evaluating this key topological property has been devised by Chern [24].

First, the original 1-form A is, in the language of Chern, "projectivized". What this means is that the original coefficients of the 1-form A are made homogenous of degree zero by division with the function $\lambda$, which is the square root of the sums of squares of the original coefficient functions. This new projectivized 1 form can also be used to compute its own Pfaff sequence, and its various electromagnetic-like fields, **E'** and **B'**. All of the preceding constructions, such as dA, A^dA, dA^dA, follow in terms of the projectivized 1-form.

Following Chern, the Euler index becomes the integral

$$X = \text{const} \iint \iint K = \text{const} \iint \iint -2 \; \mathbf{E'} \bullet \mathbf{B'} \; dx^\wedge dy^\wedge dz^\wedge dt. \qquad (19.1)$$

Note that the field intensities **E'** and **B'** are not exactly the same as the fields **E** and **B** generated by the original non-projectivized 1-form.

In Lagrangian field theories of a projectivized Action, a non-zero value for K implies that the second Chern class is not empty and signals the demise of time reversal and parity symmetry [25] (hence, the name Topological Parity 4-form). It should be remarked that K is the exterior derivative of the 3-form of topological torsion, H, and that this 3-form can be put into correspondence with the Chern-Simons 3-form of differential geometry. In effect the evolutionary law for the 3-form of Topological Torsion given by (5.7) is a Lagrangian field theory built on a Chern-Simons action. In this article, no constraint of self dualism is imposed, as is usually the case in current string theories.

When the electric field of the projectivized 1-form is orthogonal to the magnetic field, then the Euler index is zero. The idea that this Poincare invariant might have deeper meaning led Eddington [26] to state: "It is somewhat curious that the scalar-product of the electric and magnetic forces is of so little importance in classical theory, for (eq (18.7)) would seem to be the most fundamental invariant of the field. Apart from the fact that it vanishes for electromagnetic waves propagated in the absence of any bound electric field (i.e., remote from electrons), this invariant seems to have no significant properties. Perhaps it may turn out to have greater importance when the study of electron-structure is more advanced."

A non-zero value of the Topological Parity 4-form, K, implies that the divergence of T is not zero. Therefore, torsion lines can stop or start within the variety even though the evolution is C2 continuous. The torsion current is not necessarily conserved and 3-dimensional defects can be produced internally. String theorists describe this effect as an anomaly of the axial (Torsion) current. In the same sense that the closed but not exact 1-form leads to a complex representation involving ordered pair of variables, a closed but not exact 3-form leads to a quaternionic representation.

The concept of a domain of non-null Euler index ($K \neq 0$) now appears to be useful to the theory of magnetic reconnection in the electromagnetic case [27] and to vortex reconnection [28] in the hydrodynamic case. The correspondence between the bridging and rib structures produced in numerical simulations of turbulent fluid flows and the 4-string interaction of superstring theory is remarkable [29]. The concept ($K \neq 0$) appears to be applicable to the understanding of the stretching of lines and surfaces in turbulent flows where time-reversal symmetry is violated [30]. The appearance of large scale structures in certain flows has been associated with the lack of parity invariance [31]. The concepts of macroscopic violations of P and T symmetries appear to have application to the theory of the quantum Hall effect [32].

With regards to hydrodynamic systems, the evolution of a flow from a laminar flow to a turbulent flow involves topological evolution. For the Navier-Stokes system, the Euler index depends upon the viscosity and the lack of Frobenius integrability of the vorticity field (see equation 17.9). Such a term yields a local source for the creation of Torsion currents. The lack of reversibility of such flows, and the irreducible time dependent, 3 dimensional features of such flows, implies that K can not be zero for the turbulent state. It is conjectured that the Euler index of the flow (the integral of K over the domain) is not zero during the transition to turbulence. That is, K is not a last multiplier of the spatial volume element, dx^dy^dz for the flow describing the continuous (relative to the Cartan C2 topology) transition to turbulence. If dQ^F = 0 then the function K defines an integrating actor in the sense of a mass density such that

$$\text{div } (K\mathbf{V} + \partial K/\partial t = 0. \tag{19.2}$$

If K were a mass density, this equation is often called the "equation of continuity", but it is more accurately described as the "conservation of mass". Relative to the Cartan topology all C2 vector fields are continuous. The transition to the turbulent state, however, must be discontinuous, for the Cartan topology in the turbulent state is disconnected.

## 20. SUMMARY

To review, a topology has been constructed on a variety in terms of the elements of closure of a Cartan system of C2 differential forms and their intersections. The associated topological structure indicates that all processes generated by the Lie convective derivative (relative to a C2 vector field, V) are continuous relative to the Cartan topology. However, the processes so generated are not necessarily homeomorphisms for they need not be reversible; i.e., the topology of the initial state can evolve continuously into a different topology on the final state. The method for constructing the Cartan topology is the same on both the initial and the final state, but, for example, the "hole and handle" count on the initial state can be different from the "hole and handle" count in the final state.

In terms of a single 1-form of Action, A, a Cartan topological base was constructed in terms of a set of distinct elements, defined as a Pfaff sequence, and their closures. The fundamental laws of evolution of each of the elements of the topological base was formulated relative to an arbitrary vector field. It was determined that there are two categories of continuous flows, those which are "closed" and those which are "open". A special sub-category of closed flows describe a Hamiltonian evolution, an evolutionary process which preserves the number of "holes and handles".

Relative to the closed category of continuous processes, all even dimension elements of the Cartan topological base are evolutionary invariants. For closed flows, topological evolution takes place only in terms of the odd elements of the topological base. The first odd element of the topological base is the Action, and its law of evolution is equivalent to the evolution of energy. The next odd element (and the only other odd element on space-time) of the Cartan topological base is formulated as the novel 3-form of Topological Torsion. The evolution of this 3-form is studied, for although it does not necessarily satisfy a local conservation law, the anomalous source term, defined as topological parity, can be computed. It is a source of system evolutionary defects.

However, it is still possible to establish a set of global conservation laws for the category of closed, (uniformly) continuous but irreversible evolutionary flows. Although the evolution of topological torsion may be described by a continuous process, the creation of topological torsion from a state without topological torsion is not described by a continuous process. As the Cartan topology is not connected, the creation of topological torsion must involve discontinuous processes or shocks.

The fundamental equation of topological evolution, $L(\rho V)A = Q$, is equivalent to cohomological format of the first law of thermodynamics, $W+dU = Q$. The heat 1-form Q may be used to form a Pfaff sequence whose Pfaff dimension may be used to further classify evolutionary flows. For example, if the Pfaff dimension of Q is 2 or less, then Q can be written in the equilibrium format, $Q = TdS$. An example of an open system of flows (defined as $dQ \neq 0$) was presented in terms of the Navier-Stokes equations, for which the anomalous source term, can be computed. In effect it was demonstrated that C2 irreversible flows are among the solution set to the Navier-Stokes system. An abstract example was also given for an electromagnetic Action, in which the concept of time reversal and parity symmetry breaking was associated with a non-null Euler characteristic of the Cartan topology.

### 20. Acknowledgments

Although bits and pieces of the general ideas that appear in this article formed the basis of a number of research applications and publications, the entire early theory has never appeared in print. The first draft was completed about 1979 after receiving encouragement and support form NASA-AMES in 1977 in attempts to study the onset of turbulence in terms of the 3-form A^dA [33]. Generalizations of such theories are now called Chern Simons theories. A second revision was completed about 1987 following the recognition by the author and Phil Baldwin of the fact that the Cartan Topology was a disconnected topology for Pfaff dimension 3 or more [34]. The author wishes to express his appreciation to Phil Baldwin for a careful reading of the second manuscript, and for his suggestions of presentation to make the article appeal to a broader audience. About the same time (1986) the recognition of topological defects in a swimming pool (Falaco Solitons [35]) greatly stimulated the research in the area of continuous topological evolution. Following a report on the Falaco effect at Dynamics Days in Austin 1987, the comments of Dennis Sullivan led to a rewording that emphasizes the idea that the topology under consideration is the topology of the dynamical Pfaff system. Throughout the entire period, the insistence of E. J. Post on the importance of closed 3-forms in physical systems (Topological Spin [20] and Topological Torsion [16]) gave encouragement when needed. The extension of the ideas to implicit surfaces, Finsler spaces and spaces with affine torsion started about 1993, and has led to some recent advances in the understanding of the origin of charge current-potential interactions. Most of the recent advances and applications appear on the website
http://www.cartan.pair.com .

The generosity of Mike Gorman in lending to me the resources of his computing and laboratory facilities in the 80's led to the pictorial displays so helpful to the presentation of topological ideas. This work was supported in part, over the years, by the Energy Laboratory at the University of Houston.

\